\documentclass[reprint,pra,amsmath,amssymb,aps,showpacs,preprintnumbers]{revtex4-2}
\usepackage{graphicx} 
\usepackage{datetime}
\usepackage{bm} 

\begin{document}
\title{Universal optical polarizability for plasmonic nanostructures}
\author{Tigran V. Shahbazyan}
\affiliation{Department of Physics, Jackson State University, Jackson, Mississippi 39217 USA}


\begin{abstract} 
We develop an analytical model for calculation of optical spectra for metal nanostructures of arbitrary shape supporting localized surface plasmons (LSPs). For plasmonic systems with characteristic size below the diffraction limit, we obtain an explicit expression for optical polarizability that describes the lineshape of optical spectra solely in terms of the metal dielectric function and  LSP frequency. The amplitude of the LSP spectral band is determined by the effective system volume that, for long-wavelength LSPs, can significantly exceed the physical volume of metal nanostructure.  Within the quasistatic approach, we derive the exact  LSP Green's function and establish general spectral properties of LSPs,  including the distribution and oscillator strength of the LSP states. These results can be used to model or interpret the experimental spectra of plasmonic nanostructures and to tune their optical properties for various applications.
\end{abstract}
\maketitle


\section{Introduction}
 
Localized surface plasmons (LSPs) are collective electron excitation resonantly excited by incident light in metal nanostructures with  characteristic size below  the diffraction limit \cite{atwater-jap05,ozbay-science06,stockman-review}. Optical interactions between the LSPs and excitons in dye molecules or semiconductors underpin numerous phenomena in the plasmon-enhanced spectroscopy, such as surface-enhanced Raman scattering \cite{sers}, plasmon-enhanced fluorescence and luminescence \cite{feldmann-prl02,artemyev-nl02,novotny-prl06,sandoghdar-prl06,halas-nl07,halas-acsnano09,ming-nl09,pustovit-prl09}, strong exciton-plasmon coupling \cite{bellessa-prl04,sugawara-prl06,wurtz-nl07,fofang-nl08,hakala-prl09,manjavacas-nl11,salomon-prl12,garcia-prl13,antosiewicz-acsphotonics14,luca-apl14,shahbazyan-nl19}, and plasmonic laser (spaser) \cite{bergman-prl03,stockman-natphot08,noginov-nature09,shahbazyan-acsphot17}. Optical properties of metal nanostructures of various sizes and shapes are of critical importance for numerous plasmonics applications \cite{vanduyne-arpc07,zijlstra-acssens17,jin-nature20}, and were therefore extensively studied  experimentally and theoretically \cite{elsayed-jpcb99,schatz-jpcb03,noguez-jpcb03,elsayed-jpcb06,noguez-jpcc07,link-csr15,abajo-csr17}. 
The optical polarizability tensor  $\bm{\alpha}(\omega)$ of a plasmonic nanostructure determines its  response  to an incident electromagnetic (EM) field $\bm{E}_{\rm in}e^{-i\omega t}$, where $\omega$ is the incident field frequency, and, at the same time, defines the optical interactions between the LSPs and excitons. If the characteristic system size is much smaller than the radiation wavelength, so that $\bm{E}_{\rm in}$ is nearly uniform on the system scale, the induced dipole moment of a plasmonic nanostructure has the form  $\bm{p}(\omega)=\bm{\alpha}(\omega)\bm{E}_{\rm in}$, where $\bm{\alpha}(\omega)$ can be calculated, with a good accuracy, within the quasistatic approach \cite{stockman-review}.  Fully analytical models for $\bm{\alpha}(\omega)$ have long been available for systems of highly symmetric shapes, such as spherical, ellipsoidal or cylindrical structures \cite{schatz-jpcb03}. For example, a metal nanosphere of radius $a$ placed in the air is characterized by the scalar polarizability
 \begin{equation}
 \label{pol-sphere}
 \alpha(\omega)=a^{3}\, \dfrac{\varepsilon(\omega)-1}{\varepsilon(\omega)+2},
 \end{equation}
where $\varepsilon(\omega)=\varepsilon'(\omega)+i\varepsilon''(\omega)$ is  complex dielectric function of the metal. For  more complicated shapes, several  models have been suggested as well which, however, contain some parameters  to be calculated numerically \cite{schatz-jpcb03,noguez-jpcb03,elsayed-jpcb06,noguez-jpcc07,abajo-csr17}.

On the other hand, due to uncertainties in   the shape  and size  of  actual structures explored in the experiment, the analytical or numerical models describing \textit{both} the LSP frequency and the lineshape of optical spectra, as Eq.~(\ref{pol-sphere}) does, are not even necessary. Typically, the spectral position of the LSP resonance peak is measured with a reasonably high accuracy, and so the main challenge is to describe or interpret the spectral lineshape \cite{link-csr15,pini-sr22}. Here, we present an analytical model describing accurately the optical spectra of plasmonic  nanostructures of \textit{arbitrary} shape with LSP frequencies treated as \textit{input} parameters.

Specifically,  the optical polarizability tensor of a small metal nanostructure  supporting LSP resonance at a frequency $\omega_{n}$ has the form $\bm{\alpha}_{n}(\omega)=\alpha_{n}(\omega)\bm{e}_{n}\bm{e}_{n}$, where 
 \begin{equation}
 \label{pol-small}
 \alpha_{n}(\omega)=V_{n}\, \dfrac{\varepsilon(\omega)-1}{\varepsilon(\omega)-\varepsilon'(\omega_{n})},
 \end{equation}
is the scalar polarizability, $\bm{e}_{n}$ is the unit vector for  LSP mode polarization, and $V_{n}=V_{\rm m}|\chi'(\omega_{n})|s_{n}$ is the effective volume. Here, $V_{\rm m}$ is the metal volume, $\chi'(\omega)=[\varepsilon'(\omega)-1]/4\pi$ is the real part of susceptibility (we use Gaussian units), and the parameter $s_{n}\leq 1$ depends on the system geometry. Thus, for any geometry, the lineshape of optical spectra is determined only by the  metal dielectric function and  the LSP frequency, while the spectral peak amplitude depends on the system effective  volume. The polarization (\ref{pol-small}) can be extended to larger systems by including the LSP radiation damping. 

To obtain Eq.~(\ref{pol-small}), we employed the LSP Green's function approach in the quasistatic regime \cite{shahbazyan-prl16,shahbazyan-prb18,shahbazyan-prb21} (see Appendix).  Within this approach, we have also established several exact relations characterizing the distribution of LSP states.

\section{LSP Green's function}

We consider  a metal nanostructure supporting a LSP that is localized at a length scale much smaller than the radiation wavelength. In the absence of retardation effects, each region of the structure, metallic or dielectric, is characterized by the dielectric function $\varepsilon_{i}(\omega)$, so that the full dielectric function is $\varepsilon (\omega,\bm{r})=\sum_{i}\theta_{i}(\bm{r})\varepsilon_{i}(\omega)$, where $\theta_{i}(\bm{r})$ is the  unit step function that vanishes outside of the region volume $V_{i}$. We assume that dielectric regions' permittivities are constant,  and adopt $\varepsilon(\omega)$ for the metal region. The LSP modes are defined by the lossless   Gauss equation as \cite{stockman-review}, 
%
\begin{equation}
\label{gauss-law}
\bm{\nabla}\cdot\left [\varepsilon' (\omega_{n},\bm{r})\bm{\nabla} \Phi_{n}(\bm{r})\right ]=0,
\end{equation}
where $\Phi_{n}(\bm{r})$ and $\bm{E}_{n}(\bm{r})=-\bm{\nabla} \Phi_{n}(\bm{r})$ are the mode's potential and electric field, which we chose real. Note that the eigenmodes of Eq.~(\ref{gauss-law}) are orthogonal in  each  region \cite{shahbazyan-prb21}:
$\int\! dV_{i} \bm{E}_{n}(\bm{r})\!\cdot\!\bm{E}_{n'}(\bm{r})=\delta_{nn'}\int\! dV_{i} \bm{E}_{n}^{2}(\bm{r})$.

The EM dyadic Green's function $\bm{D} (\omega;\bm{r},\bm{r}') $ satisfies (in the operator form) $\bm{\nabla}\!\times\! \bm{\nabla}\!\times \bm{D}-(\omega^{2}/c^{2})\varepsilon \bm{D} =(4\pi\omega^{2}/c^{2})\bm{I}$, where $\bm{I}$ is the unit tensor, while the longitudinal part  of $\bm{D}$ is obtained by applying the operator $\bm{\nabla}$ to both sides. In the near field, we switch to the scalar Green's function for the potentials $D(\omega;\bm{r},\bm{r}')$, defined as $\bm{D} (\omega;\bm{r},\bm{r}')=\bm{\nabla}\bm{\nabla}'D(\omega;\bm{r},\bm{r}')$, which satisfies [compare to Eq.~(\ref{gauss-law})]
\begin{equation}
\label{gauss-green-pot}
\bm{\nabla}\cdot\left [\varepsilon (\omega,\bm{r})\bm{\nabla}D(\omega;\bm{r},\bm{r}')\right ]=4\pi \delta(\bm{r}-\bm{r}').
\end{equation}
We now adopt the decomposition $D=D_{0}+D_{\rm LSP}$, where $D_{0}(\bm{r}-\bm{r}')=-|\bm{r}-\bm{r}'|^{-1}$ is the free-space Green's function and $D_{\rm LSP}(\omega;\bm{r},\bm{r}')$ is the LSP contribution. The latter is expanded over the eigenmodes  of Eq.~(\ref{gauss-law}) as \cite{shahbazyan-prl16,shahbazyan-prb18,shahbazyan-prb21} (see Appendix)
\begin{equation}
\label{green-exp}
D_{\rm LSP}(\omega;\bm{r},\bm{r}')=\sum_{n}D_{n}(\omega)\Phi_{n}(\bm{r})\Phi_{n}(\bm{r}'),
\end{equation}
where the coefficients $D_{n}(\omega)$ have the form
\begin{equation}
\label{mode-coeff}
D_{n}(\omega)= 
\dfrac{4\pi}{\int\! dV \bm{E}_{n}^{2}(\bm{r})} 
-\dfrac{4\pi}{\int\! dV \varepsilon (\omega,\bm{r})\bm{E}_{n}^{2}(\bm{r})}.
\end{equation}
The first term in Eq.~(\ref{mode-coeff}) ensures the boundary condition for $\varepsilon=1$ and will be omitted in the following. While the expansion in Eq.~(\ref{green-exp}) runs over the eigenmodes of the lossless Gauss equation (\ref{gauss-law}), the coefficients $D_{n}$ depend on  complex  $\varepsilon(\omega,\bm{r})=\varepsilon'(\omega,\bm{r})+i\varepsilon''(\omega,\bm{r})$ (see Appendix).
Accordingly, the LSP dyadic Green's function for the electric fields has the form $\bm{D}_{\rm LSP} (\omega;\bm{r},\bm{r}')=\sum_{n}D_{n}(\omega)\bm{E}_{n}(\bm{r})\bm{E}_{n}(\bm{r}')$.

We now  note that, in the quasistatic regime, the  frequency and coordinate dependencies in the LSP Green's function can be separated out. Using the Gauss equation (\ref{gauss-law}) in the integral form $\int\! dV \varepsilon' (\omega_{n},\bm{r})\bm{E}_{n}^{2}(\bm{r})=0$, the volume  integral in Eq.~(\ref{mode-coeff})  can be presented as 
\begin{align}
\int\! dV \varepsilon (\omega,\bm{r})\bm{E}_{n}^{2}(\bm{r})
=\left [\varepsilon (\omega)-\varepsilon' (\omega_{n})\right ] \int\! dV_{\rm m}\bm{E}_{n}^{2}(\bm{r}),
\end{align}
where integration in the right-hand side is  carried over the \textit{metal} volume $V_{\rm m}$, while  the dielectric regions' contributions,  characterized by constant permittivities,  cancel each other out. The LSP Green's function takes the form
\begin{equation}
\label{lsp-green}
\bm{D}_{\rm LSP} (\omega;\bm{r},\bm{r}')=-\sum_{n}\dfrac{4\pi}{\int\! dV_{\rm m} \bm{E}_{n}^{2}}\frac{\bm{E}_{n}(\bm{r})\bm{E}_{n}(\bm{r}')}{\varepsilon (\omega)-\varepsilon' (\omega_{n})},
\end{equation}
which represents the basis for our further analysis of the optical properties of metal nanostructures. Note that 
near the LSP pole, the denominator of Eq.~(\ref{lsp-green}) can be expanded as $\varepsilon (\omega)-\varepsilon' (\omega_{n})=[\partial\varepsilon' (\omega_{n})/\partial \omega_{n}](\omega-\omega_{n}+i\gamma_{n}/2)$, where $\gamma_{n}=2\varepsilon'' (\omega_{n})/[\partial\varepsilon' (\omega_{n})/\partial \omega_{n}]$ is the LSP  decay rate \cite{stockman-review}, and we recover the Lorentzian approximation for the LSP Green's function \cite{shahbazyan-prl16,shahbazyan-prb18,shahbazyan-prb21}.

\section {LDOS, DOS, and mode volume}

Using representation (\ref{lsp-green}) for the LSP Green's function, we can establish some general spectral properties of LSPs. In the following, we consider metal nanostructures of arbitrary shape in a dielectric medium with permittivity $\varepsilon_{d}$ (we set $\varepsilon_{d}=1$ for now). We assume that $\omega$ lies in the plasmonics frequency domain, i.e., $|\varepsilon''(\omega)/\varepsilon'(\omega)|\ll 1$, and so the LSP quality factor $Q_{n}=\omega_{n}/\gamma_{n}=\omega_{n}[\partial\varepsilon' (\omega_{n})/\partial \omega_{n}]/2\varepsilon'' (\omega_{n})$ is sufficiently large \cite{stockman-review}. An important quantity that is critical in many applications is the local density of states (LDOS), which describes  the number of LSP states in the unit volume and  frequency interval:
\begin{equation}
\rho(\omega,\bm{r})=\dfrac{1}{2\pi^{2}\omega}\text{Im}\,\text{Tr}\bm{D}_{\rm LSP} (\omega;\bm{r},\bm{r})=\sum_{n}\rho_{n}(\omega,\bm{r}).
\end{equation}
Here, $\rho_{n}(\omega,\bm{r})$ is the LDOS for an individual LSP mode which, using the Green's function (\ref{lsp-green}), takes the form
\begin{equation}
\label{ldos}
\rho_{n}(\omega,\bm{r})
=\dfrac{2}{\pi\omega}\dfrac{\bm{E}_{n}^{2}(\bm{r})}{\int\! dV_{\rm m} \bm{E}_{n}^{2}}\,\text{Im}\!\left [\frac{-1}{\varepsilon (\omega)-\varepsilon' (\omega_{n})}\right ].
\end{equation}
Integration of the LDOS over the volume yields the LSP density of states (DOS) $\rho_{n}(\omega)=\int\! dV\rho_{n}(\omega,\bm{r})$, describing the number of LSP states per unit frequency interval. To elucidate the distribution of LSP states  in the system, let us compare the LSP  DOS inside the metal, $\rho_{n}^{\rm m}(\omega)=\int\! dV_{\rm m}\rho_{n}(\omega,\bm{r})$, and in the surrounding dielectric medium, $\rho_{n}^{\rm d}(\omega)=\int\! dV_{\rm d}\rho_{n}(\omega,\bm{r})$. From Eq.~(\ref{ldos}), $\rho_{n}^{\rm m}(\omega)$ is readily obtained as
\begin{equation}
\label{dos-in}
\rho_{n}^{\rm m}(\omega)
=\dfrac{2}{\pi\omega}\,\text{Im}\!\left [\frac{-1}{\varepsilon (\omega)-\varepsilon' (\omega_{n})}\right ].
\end{equation}
To evaluate $\rho_{n}^{\rm d}(\omega)$, we use the Gauss equation to present the integral over the dielectric region outside the metal as $\int\! dV_{\rm d}\bm{E}_{n}^{2}=-\varepsilon' (\omega_{n})\int\! dV_{\rm m} \bm{E}_{n}^{2}$, yielding 
\begin{equation}
\label{dos-out}
\rho_{n}^{\rm d}(\omega)
=\dfrac{2}{\pi\omega}\,\text{Im}\!\left [\frac{\varepsilon' (\omega_{n})}{\varepsilon (\omega)-\varepsilon' (\omega_{n})}\right ].
\end{equation}
Since for typical LSP frequencies $|\varepsilon' (\omega_{n})|\gg 1$, we have $\rho_{n}^{\rm d}(\omega)=|\varepsilon' (\omega_{n})|\rho_{n}^{\rm m}(\omega)\gg \rho_{n}^{\rm m}(\omega)$,
implying that the LSP states are primarily distributed \textit{outside} the metal. The full LSP DOS $\rho_{n}(\omega)=\rho_{n}^{\rm m}(\omega)+\rho_{n}^{\rm d}(\omega)$ has the form
\begin{equation}
\label{dos}
\rho_{n}(\omega)
=\dfrac{2}{\pi\omega}\,\text{Im}\!\left [\frac{\varepsilon' (\omega_{n})-1}{\varepsilon (\omega)-\varepsilon' (\omega_{n})}\right ],
\end{equation}
which is valid for \textit{any}  nanostructure shape. 

Let us now evaluate the number of LSP states per mode, $N_{n}=\int\!d\omega\rho_{n}(\omega)$. Performing the frequency integration in the Lorentzian approximation,  we obtain
\begin{equation}
\label{mode-number}
N_{n}=\dfrac{2|\varepsilon' (\omega_{n})-1|}{\omega_{n} \partial\varepsilon' (\omega_{n})/\partial \omega_{n}}.
\end{equation}
For the Drude form of $\varepsilon (\omega)$, Eq.~(\ref{mode-number}) yields $N_{n}\approx 1$, implying that the LSP states saturate the mode's oscillator strength. However, for the  experimental dielectric function, $N_{n}$ can be substantially below its maximal value, which has implications for the  optical spectra (see below).

Another important quantity that characterizes the local field confinement  is  the LSP mode volume ${\cal V}_{n}$, which is related to the  LDOS as  ${\cal V}_{n}^{-1}=\rho_{n}(\bm{r})=\int d\omega\rho_{n}(\omega,\bm{r})$, where $\rho_{n}(\bm{r})$ is the LSP spatial density \cite{shahbazyan-prl16,shahbazyan-prb18}. Performing the frequency integration, we obtain 
\begin{equation}
\label{mode-volume}
\frac{1}{{\cal V}_{n}}=\int d\omega\rho_{n}(\omega,\bm{r})=
\dfrac{2\bm{E}_{n}^{2}(\bm{r})}{[\omega_{n} \partial\varepsilon' (\omega_{n})/\partial \omega_{n}]\!\int\! dV_{\rm m} \bm{E}_{n}^{2}}.
\end{equation}
While  the LSP mode volume is a \textit{local} quantity that can be very small  [i.e., the density $\rho_{n}(\bm{r})$ can be very large] at hot spots,  its integral is  bound as $\int\!dV/{\cal V}_{n}=N_{n}\leq 1$. 

\section{Optical polarizability}

Consider now a metal nanostructure  in the incident EM field $\bm{E}_{\rm in}e^{-i\omega t}$ that is nearly uniform on the system scale. The  system's induced dipole moment is obtained by volume integration  of  the electric polarization vector, $\bm{p}(\omega)=\chi(\omega)\int dV_{\rm m} \bm{E}(\omega,\bm{r})$, where $\bm{E}(\omega,\bm{r})$ is the local field inside the metal, given by
\begin{equation}
\label{field-full}
\bm{E}(\omega,\bm{r})=\bm{E}_{\rm in}+ \chi(\omega) \int \! dV'_{\rm m}  \bm{D}_{\rm LSP}(\omega;\bm{r},\bm{r}')\bm{E}_{\rm in}.
\end{equation}
Using the LSP Green's function (\ref{lsp-green}), we obtain
\begin{equation}
\label{field-full2}
\bm{E}(\omega,\bm{r})=\bm{E}_{\rm in}- 
\sum_{n} c_{n} \bm{E}_{n}(\bm{r}) \,\frac{\varepsilon (\omega)-1}{\varepsilon (\omega)-\varepsilon' (\omega_{n})},
\end{equation}
where  the coefficient $c_{n}$ is given by
\begin{equation}
\label{c-coeff}
c_{n}= \dfrac{\int\! dV_{\rm m}\bm{E}_{n}\!\cdot\!\bm{E}_{\rm in}}{\int\! dV_{\rm m} \bm{E}_{n}^{2}}.
\end{equation}
Expanding the incident field $\bm{E}_{\rm in}$  in Eq.~(\ref{field-full2}) over the LSP eigenmodes as $\bm{E}_{\rm in}=\sum_{n}c_{n}\bm{E}_{n}(\bm{r})$,  we obtain the local field inside the metal as
\begin{equation}
\label{field-inside}
\bm{E}(\omega,\bm{r})=-
\sum_{n} c_{n}\bm{E}_{n}(\bm{r}) 
\, \frac{\varepsilon' (\omega_{n})-1}{\varepsilon (\omega)-\varepsilon' (\omega_{n})}.
\end{equation}
Integrating Eq.~(\ref{field-inside}) over the system volume, multiplying the result by  $\chi(\omega)=[\varepsilon (\omega)-1]/4\pi$, and using Eq.~(\ref{c-coeff}), we obtain the plasmonic system's induced dipole moment as $\bm{p}(\omega)= \sum_{n}\bm{\alpha}_{n}(\omega)\bm{E}_{\rm in}$, where
\begin{align}
\label{polar-mode}
\bm{\alpha}_{n}= |\chi'(\omega_{n})|
 \dfrac{(\int\! dV_{\rm m}\bm{E}_{n})(\int\! dV_{\rm m}\bm{E}_{n}) }{\int\! dV_{\rm m} \bm{E}_{n}^{2}} \, \frac{\varepsilon (\omega)-1}{\varepsilon (\omega)-\varepsilon' (\omega_{n})}
\end{align}
is the LSP mode polarizability tensor  [here, $\varepsilon' (\omega_{n})-1=-4\pi |\chi'(\omega_{n})|$]. We now introduce the LSP mode polarization unit vector  as  $\bm{e}_{n}=\int\! dV_{\rm m}\bm{E}_{n}/|\int\! dV_{\rm m}\bm{E}_{n}|$ and  the \textit{effective} system volume $V_{n}$ as
\begin{equation}
\label{Vn}
V_{n}=V_{\rm m}|\chi'(\omega_{n})|s_{n},
~~~
s_{n}=\dfrac{\left (\int\! dV_{\rm m}\bm{E}_{n}\right )^{2}}{V_{\rm m}\int\! dV_{\rm m} \bm{E}_{n}^{2}}.
\end{equation}
Then, using  Eqs.~(\ref{polar-mode}) and (\ref{Vn}), we obtain the polarizability tensor $\bm{\alpha}_{n}(\omega)=\alpha_{n}(\omega)\bm{e}_{n}\bm{e}_{n}$, where the scalar polarizability  $\alpha_{n}(\omega)$  is given by Eq.~(\ref{pol-small}).

The parameter $s_{n}$ in the effective volume  (\ref{Vn}) depends on the system geometry and characterizes the strength of LSP coupling  to the external EM field. Namely, it describes the relative variation of the LSP mode field inside the metal  structure, while being independent of its overall amplitude. For the dipole LSP modes, which have no nodes  inside the nanostructure, $s_{n}$ is nearly independent of the metal volume. For nanoparticles of spherical or spheroidal shape, its exact value is $s_{n}=1$ (see Appendix), while smaller values $s_{n}\lesssim 1$ are expected for other geometries. For higher-order LSP modes, whose electric fields oscillate inside the structure and, hence, have small overlap with the incident field,  the parameter $s_{n}$ is small. 

The polarizability (\ref{pol-small}) is valid for small nanostructures characterized by weak LSP radiation damping as compared to the Ohmic losses in metal. For larger systems, to satisfy the optical theorem, the LSP radiation damping  must be included by considering the system's interaction with the radiation field, which leads to the replacement  $\alpha_{n} \rightarrow \alpha_{n} [1-(2i\omega^{3}/3c^{3})\alpha_{n}]^{-1}$, where $c$ is the speed of light \cite{carminati-oc06,novotny-book}. For such systems, after restoring the permittivity of surrounding medium $\varepsilon_{d}$, the  scalar polarizability  takes the form
 \begin{equation}
 \label{pol-rad}
\alpha_{n}(\omega)
 =V_{n}\, \dfrac{\varepsilon(\omega)-\varepsilon_{d}}{\varepsilon(\omega)-\varepsilon'(\omega_{n})-\frac{2i}{3}k^{3}V_{n}[\varepsilon(\omega)-\varepsilon_{d}]},
 \end{equation}
where $k=\sqrt{\varepsilon_{d}}\omega/c$ is the light wave vector, while the system effective volume is now given by
\begin{equation}
\label{volume-eff}
V_{n}=V_{\rm m}|\varepsilon'(\omega_{n})/\varepsilon_{d}-1|s_{n}/4\pi.
\end{equation}
The optical polarizability (\ref{pol-rad}) is the central result of this Letter which permits  accurate description of optical spectra for diverse plasmonic structures, including those of irregular shape, using, as input,  only the basic system parameters and the  LSP frequency. 
In terms of $\alpha_{n}$,   the extinction and scattering cross-sections  have the form \cite{novotny-book}
\begin{equation}
\label{cross}
\sigma_{\rm ext}(\omega)=\frac{4\pi \omega}{c}|\epsilon_{n}|^{2}\alpha''_{n}(\omega),
~
\sigma_{\rm sc}(\omega)=\frac{8\pi \omega^{4}}{3c^{4}}|\epsilon_{n}|^{2}\left |\alpha_{n}(\omega)\right |^{2},
\end{equation}
where $\epsilon_{n}=\bm{e}_{n}\cdot\bm{E}_{\rm in}/|\bm{E}_{\rm in}|$ is  the LSP polarization relative to the incident light. 

Note that Eq.~(\ref{pol-rad}) reproduces the  known analytical results for nanostructures of simple shapes. For  a nanosphere of radius $a$, we have $s_{n}=1$, $\varepsilon'(\omega_{n})=-2$, and we recover Eq.~(\ref{pol-sphere}) with the effective volume $V_{n}=a^{3}$, which is significantly smaller than the system volume. The polarizability (\ref{pol-rad}) also matches the known result for spheroidal nanoparticles (see Appendix).
For metal structures with multiple LSP resonances, including porous structures \cite{klar-nl18}, the polarizability tensor is $\bm{\alpha}(\omega)=\sum_{n}\alpha_{n}(\omega)\bm{e}_{n}\bm{e}_{n}$, where $V_{n}$ can  now be considered as fitting parameters.

Finally, the universal form (\ref{pol-rad}) for the optical polarizability is valid for metal nanostructures embedded in dielectric medium. For more complex layered systems, including core-shell structures, the corresponding expressions for polarizability are more cumbersome and, importantly, no longer universal, i.e.,  the lineshape of optical spectra now depends explicitly (not just via the LSP frequency) on the system geometry.

  \begin{figure}[tb]
  \begin{center}
\includegraphics[width=0.99\columnwidth]{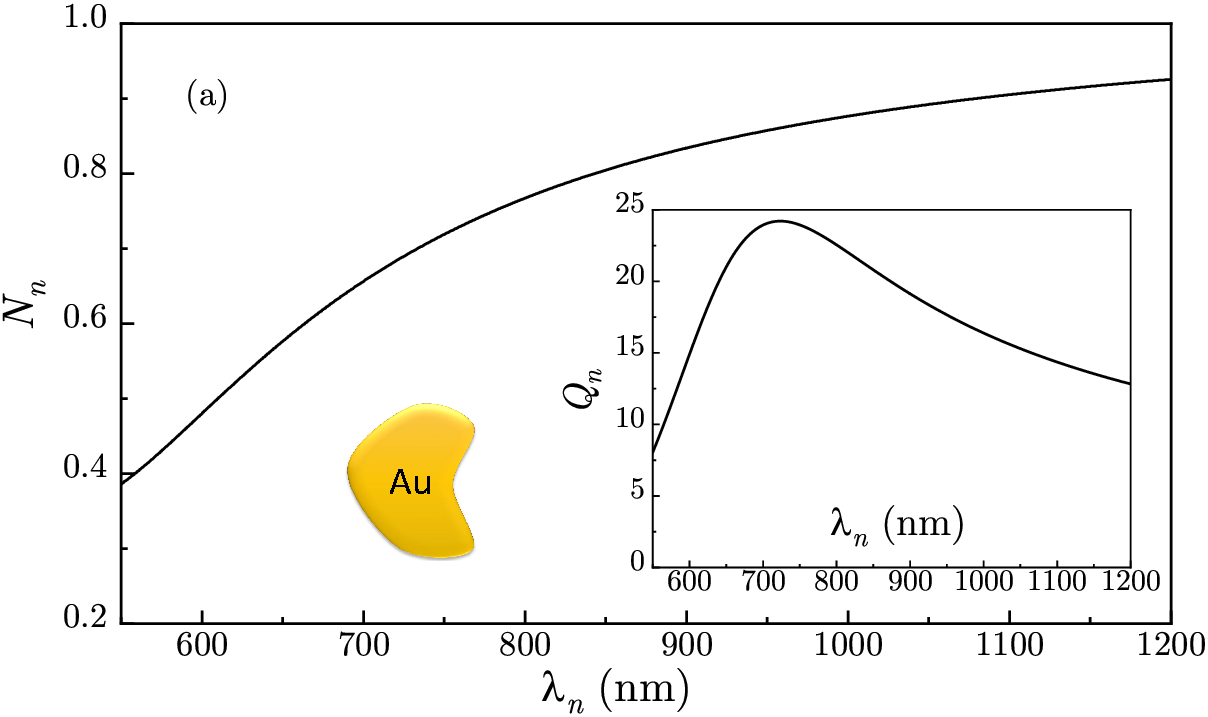}

\vspace{2mm}

\includegraphics[width=0.99\columnwidth]{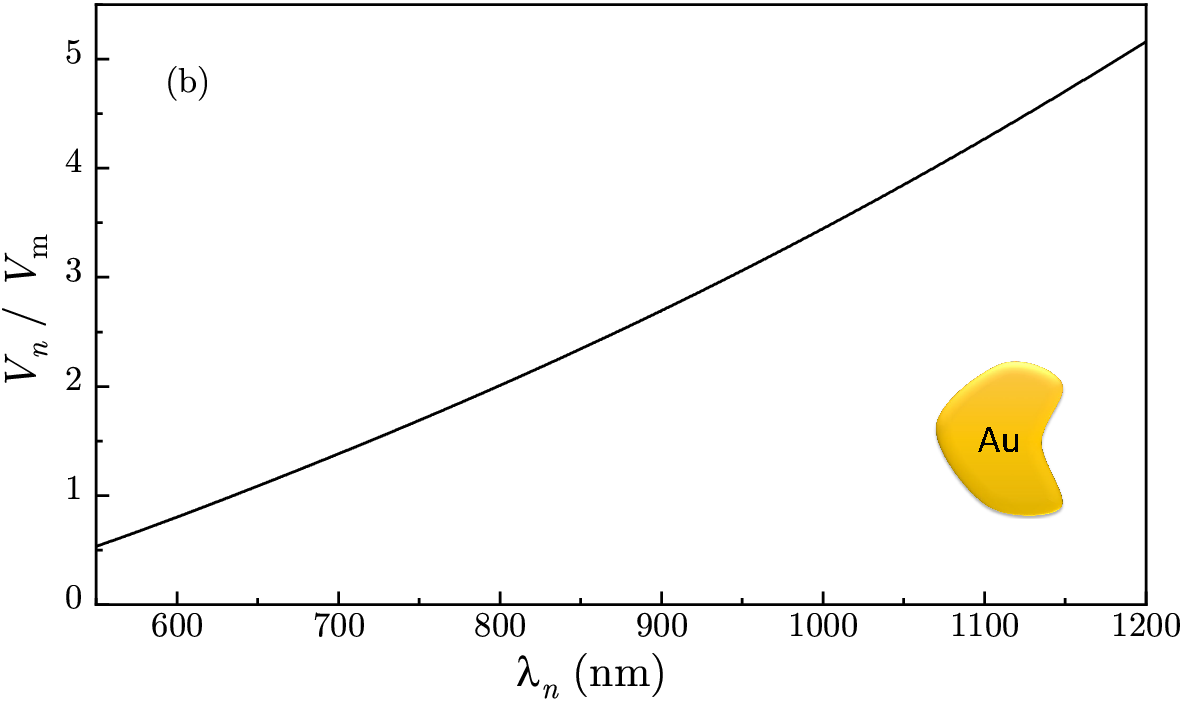}
 \end{center}
\vspace{-5mm}
\caption{\label{fig1} (a) The number of LSP states for Au  nanostructures is plotted against the LSP wavelength. Inset: the LSP quality factor wavelength dependence. (b) The normalized effective volume is plotted against the LSP wavelength.}
\vspace{-5mm}
  \end{figure}

  \begin{figure}[tb]
  \begin{center}
\includegraphics[width=0.99\columnwidth]{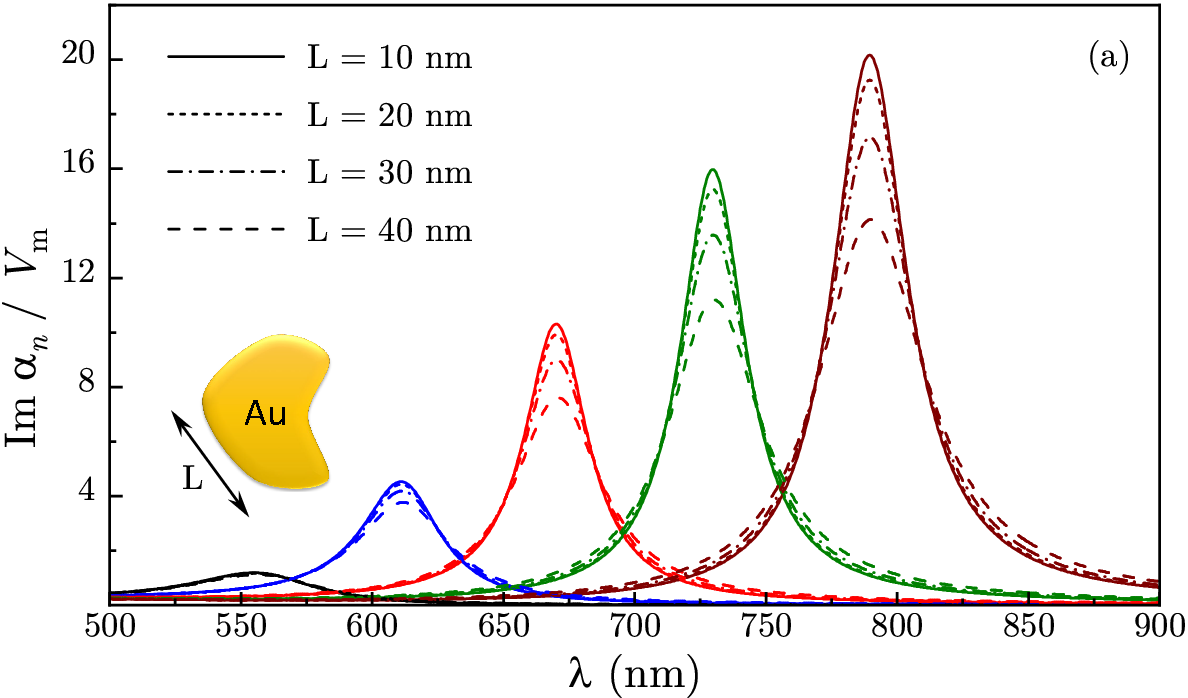}

\vspace{2mm}

\includegraphics[width=0.99\columnwidth]{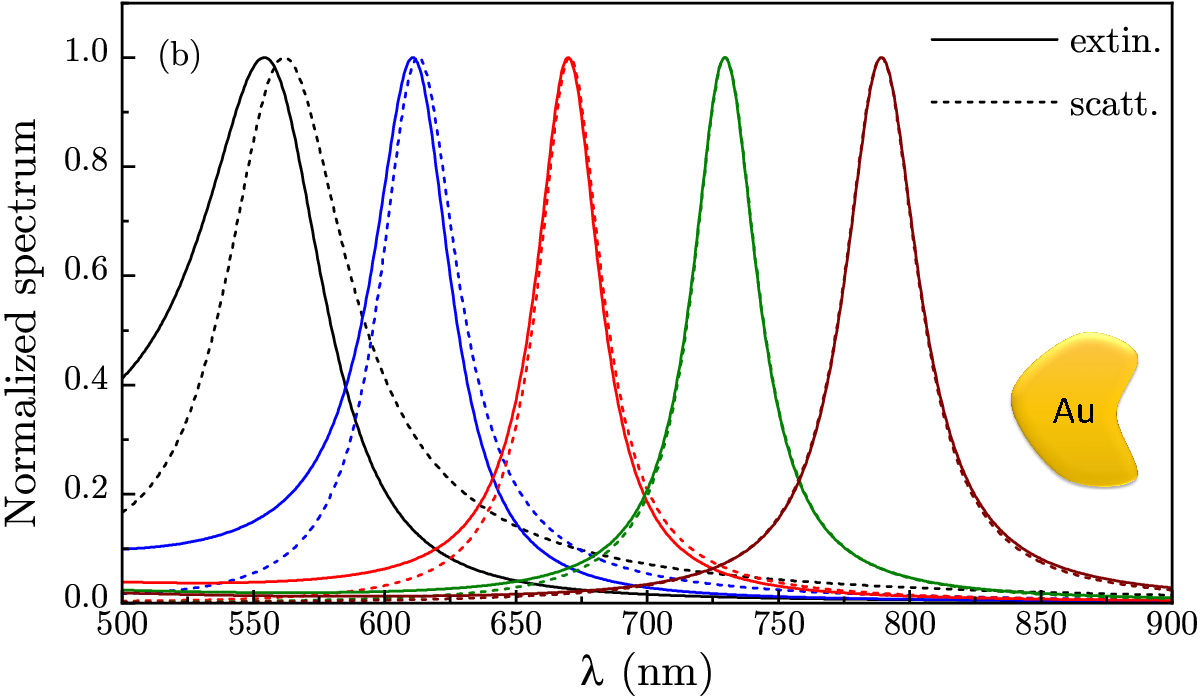}
  \end{center}
\vspace{-5mm}
\caption{\label{fig2} (a) The imaginary part of polarizability for various Au  structures is shown at different LSP wavelengths. (b) The normalized extinction and scattering spectra are shown for $L=30$ nm structures.}
\vspace{-4mm}
  \end{figure}

\section{Numerical results}

Below we present the results of numerical calculations for small gold nanostructures  to illustrate some general features of the LSP optical spectra that are common for any system geometry (we use the experimental gold dielectric function and set $s_{n}=1$). In Fig.~\ref{fig1}, we plot the number of LSP states per mode $N_{n}$ and the effective volume $V_{n}$ against the LSP wavelength $\lambda_{n}$ in the interval from 550   to 1200 nm, i.e.,  for energies below the interband transitions onset in gold. With increasing $\lambda_{n}$, as the system enters the Drude regime, $N_{n}$ increases, albeit slowly, towards its maximal value  [see Fig.~\ref{fig1}(a)]. However,  for typical LSP wavelengths from 550  to 800 nm, $N_{n}$ remains substantially below its maximal value, implying the important role of interband transitions even for energies well below the onset. Notably, $N_{n}$ does \textit{not} follow  the LSP quality factor  $Q_{n}$, shown in the inset, which peaks at $\lambda_{n}\approx 700$ nm due to the minimum of $\varepsilon''$ for gold at this wavelength.

To elucidate the effect of system geometry, in Fig.~\ref{fig1}(b), we plot the effective volume $V_{n}$ normalized by the metal volume $V_{\rm m}$ in the same LSP wavelength interval. The normalized  effective volume increases about \textit{tenfold}  from $\lambda_{n}= 550$ nm, roughly corresponding to the LSP wavelength in the gold nanosphere, to $\lambda_{n}= 1200$ nm, typical for LSPs in elongated particles with large aspect ratio. Since $V_{n}/V_{\rm m}\approx |\chi' (\omega_{n})|$, this implies that, for nanostructures of different shape but the \textit{same} metal volume, both the lineshape and peak amplitude of the optical spectra are determined by the LSP resonance position.   

In Fig.~\ref{fig2}, we show the optical spectra of gold nanostructures in water ($\varepsilon_{d}=1.77$) for different values of characteristic size $L$ and, accordingly, of metal volume $V_{\rm m}=L^{3}$, calculated using  Eqs.~(\ref{pol-rad})-(\ref{cross}) at the LSP wavelength values  550, 610, 670, 730, and 790 nm.  The imaginary part of polarizability  normalized by the metal volume increases sharply with the LSP wavelength [see Fig.~\ref{fig2}(a)], consistent with the effective volume increase in Fig.~\ref{fig1}(b). For larger structures, the LSP peak amplitudes of $\alpha_{n}''(\omega)/V_{\rm m}$ drop due to  the radiation damping. Although for full $\alpha_{n}''(\omega)$ such a decrease would be  masked by larger $V_{\rm m}$ values, it is clear that, for the \textit{same} metal volume,  radiation damping is stronger for long-wavelength LSPs since it is determined by the effective volume $V_{n}$ [see Eq.~(\ref{pol-rad})].

In Fig.~\ref{fig2}(b), we plot the extinction and scattering spectra, normalized by their respective maxima, for $L=30$ nm gold nanostructures calculated for the same LSP wavelengths as in Fig.~\ref{fig2}(a). For shorter wavelengths ($<700$ nm), the scattering spectra exhibit \textit{apparent} redshift relative to the extinction spectra. Note that, for such system size, the extinction is dominated by the absorption, implying the prominent role of non-LSP excitations in this frequency region. This behavior is consistent with a relatively low fraction (about 50\% at such wavelengths) of the LSP states per mode [see Fig.~\ref{fig1}(a)]. In the Drude regime (larger LSP wavelengths), the difference between the extinction and scattering spectra disappears as the LSP states saturate the oscillator strength.

In summary, we have developed an analytical model for optical polarization of plasmonic nanostructures of arbitrary shape whose characteristic size is below the diffraction limit. For such systems, the lineshape of optical spectra is determined by the metal dielectric function and LSP frequency while their amplitude depends on the system effective volume that increases with the LSP wavelength. We have also established some general spectral properties of the LSPs  valid for any system geometry.

\acknowledgments

This work was supported  by NSF Grants  No. DMR-2000170,  No. DMR-1856515,  and  No. DMR-1826886.

\appendix
\section{The quasistatic eigenmodes and Green function for localized surface plasmons}
\subsection{The quasistatic eigenmodes}

Here, we outline the derivation of the localized surface plasmon (LSP) Green function using the approach developed in   \cite{shahbazyan-prl16,shahbazyan-prb18,shahbazyan-prb21}. We consider  metal nanostructures supporting LSPs that are localized at the length scale much smaller than the radiation wavelength. In the absence of retardation effects, each region of the structure, metallic or dielectric, of volume $V_{i}$ is characterized by the dielectric function $\varepsilon_{i}(\omega)$, so that the full dielectric function is $\varepsilon (\omega,\bm{r})=\sum_{i}\theta_{i}(\bm{r})\varepsilon_{i}(\omega)$, where $\theta_{i}(\bm{r})$ is  unit step function that vanishes outside the region volume $V_{i}$. We assume that dielectric regions' permittivities are constant  and that only metal dielectric function $\varepsilon(\omega)$ is dispersive. The LSP modes are defined by the \textit{lossless}   Gauss's equation as \cite{stockman-review}
%
\begin{equation}
\label{app-gauss-law}
\bm{\nabla}\cdot\left [\varepsilon' (\omega_{n},\bm{r})\bm{\nabla} \Phi_{n}(\bm{r})\right ]=0,
\end{equation}
where $\Phi_{n}(\bm{r})$ and $\bm{E}_{n}(\bm{r})=-\bm{\nabla} \Phi_{n}(\bm{r})$ are the mode's potential and electric field, which we chose real. Note that the eigenmodes of Eq.~(\ref{app-gauss-law}) are orthogonal in \textit{each} region of the metal-dielectric structure (see Appendix of Ref. \cite{shahbazyan-prb21}):
%
\begin{equation}
\label{app-mode-orth}
\int\! dV_{i} \bm{E}_{n}(\bm{r})\!\cdot\!\bm{E}_{n'}(\bm{r})=\delta_{nn'}\int\! dV_{i} \bm{E}_{n}^{2}(\bm{r}).
\end{equation}
From Eq.~(\ref{app-gauss-law}), the LSP eigenmodes satisfy integral equation
\begin{equation}
\Phi_{n}(\bm{r})=-\int dV' D_{0}(\bm{r}-\bm{r}')\bm{\nabla}' \left [\chi' (\omega_{n},\bm{r}')\bm{\nabla}' \Phi_{n}(\bm{r}')\right ]
\end{equation}
where $D_{0}(\bm{r})=-1/r$ is the Green function for Laplace equation satisfying $\Delta D_{0}(\bm{r})= 4\pi\delta(\bm{r})$, and $\chi(\omega,\bm{r})=\left [\varepsilon (\omega,\bm{r})-1\right ]/4\pi$ is susceptibility. Integrating by parts and applying gradient operator $\bm{\nabla}$ to the result, one obtains
\begin{equation}
\label{app-relation-electric}
\bm{E}_{n}(\bm{r})=\int dV' \bm{D}_{0}(\bm{r}-\bm{r}') \chi' (\omega_{n},\bm{r}')\bm{E}_{n}(\bm{r}')
\end{equation}
where $\bm{D}_{0} (\bm{r}-\bm{r}')=\bm{\nabla}\bm{\nabla}'D_{0}(\bm{r}-\bm{r}')$ is  near field dyadic  Green function satisfying $\bm{\nabla}\bm{D}_{0} (\bm{r}-\bm{r}')=4\pi\bm{\nabla}'\delta(\bm{r}-\bm{r}')$. The above equation (\ref{app-relation-electric}) allows one to relate the electric field eigenmodes at any point to those inside the metal nanostructure. In particular, for a metal nanoparticle (NP) in a dielectric medium with dielectric constant $\varepsilon_{d}$ (we set $\varepsilon_{d}=1$ for now), the susceptibility $\chi'(\omega_{n},\bm{r}')$  vanishes outside the metal and one obtains
\begin{equation}
\label{app-relation-in-out}
4\pi\bm{E}_{n}(\bm{r})=\left [\varepsilon' (\omega_{n})-1\right ]\int dV_{\rm m} \bm{D}_{0}(\bm{r}-\bm{r}_{\rm m}) \bm{E}_{n}(\bm{r}_{\rm m}),
\end{equation}
where integration takes place over the metal volume $V_{\rm m}$.

The quasistatic eigenmodes $\Phi_{n}(\bm{r})$, being eigenstates of Laplace equation, represent a complete basis set, so the free-space Green function $\bm{D}_{0} (\bm{r}-\bm{r}')$ can, in principle, be expanded over the products $\Phi_{n}^{(1)}(\bm{r})\Phi_{n}^{(2)}(\bm{r}')$ of two independent eigenmodes, one of which is regular at the origin (typically inside the metal), and the other one decays for large $\bm{r}$ (outside of the metal). In the case of simple NP geometries, e.g., spherical or ellipsoidal, such expansions are known explicitly, but for arbitrary NP shape, no such an expansion is known for general positions $\bm{r}$ and $\bm{r}'$. However, in the case when $\bm{r}$ lies \textit{outside} the NP while $\bm{r}'$ lies \textit{inside} of it, such en expansion can be found in general form as $\bm{D}_{0} (\bm{r}-\bm{r}')=\sum_{n}d_{0n}\bm{E}_{n}(\bm{r})\bm{E}_{n}(\bm{r}')$, where the coefficient $d_{0n}$ is determined by inserting this expansion into Eq.~(\ref{app-relation-in-out}). We obtain
\begin{equation}
\label{app-green-free-expansion}
\bm{D}_{0} (\bm{r}-\bm{r}')=\sum_{n}\dfrac{\bm{E}_{n}(\bm{r})\bm{E}_{n}(\bm{r}')}{\int\! dV_{\rm m} \bm{E}_{n}^{2}}\frac{4\pi}{\varepsilon' (\omega_{n})-1},
\end{equation}
where $\bm{E}_{n}(\bm{r})$ and $\bm{E}_{n}(\bm{r}')$ are, respectively, the regular (inside the metal) and decaying (outside the metal) eigenmodes. It is easy to check that Eq.~(\ref{app-green-free-expansion}) reproduces accurately the known expressions for spherical and spheroidal NPs with eigenmodes $\bm{E}_{n}(\bm{r})$ expressed in terms of spherical harmonics.

\subsection{The LSP Green function}

The EM dyadic Green function $\bm{D} (\omega;\bm{r},\bm{r}') $ satisfies 
\begin{widetext}
\begin{equation}
\label{app-Green-EM}
\bm{\nabla} \times\! \bm{\nabla} \times \bm{D} (\omega;\bm{r},\bm{r}')-\frac{\omega^{2}}{c^{2}}\varepsilon (\omega,\bm{r})\bm{D} (\omega;\bm{r},\bm{r}') =\frac{4\pi\omega^{2}}{c^{2}}\bm{I}\delta (\bm{r}-\bm{r}'),
\end{equation}
\end{widetext}
where $\bm{I}$ is the unit tensor and $\delta (\bm{r}-\bm{r}')$ is the Dirac delta function. The  equation for the longitudinal part of $\bm{D}$ is obtained by applying $\bm{\nabla}$ to both sides. Since we are interested in  the near-field Green function, it is more convenient to switch, for a moment, to the scalar Green function $D(\omega;\bm{r},\bm{r}')$ for the potentials, which is defined as $\bm{D} (\omega;\bm{r},\bm{r}')=\bm{\nabla}\bm{\nabla}'D(\omega;\bm{r},\bm{r}')$, that satisfies the following equation
\begin{equation}
\label{app-gauss-green-pot}
\bm{\nabla}\cdot\left [\varepsilon (\omega,\bm{r})\bm{\nabla}D(\omega;\bm{r},\bm{r}')\right ]=4\pi \delta(\bm{r}-\bm{r}').
\end{equation}
We now adopt the decomposition $D=D_{0}+D_{\rm LSP}$, where $D_{0}=-|\bm{r}-\bm{r}'|^{-1}$ is the free-space near-field Green function and $D_{\rm LSP}$ is the LSP contribution, satisfying
\begin{widetext}
\begin{align}
\label{app-gauss-green-plas}
\bm{\nabla}\!\cdot\!\bigl[\varepsilon (\omega,\bm{r})\bm{\nabla}
D_{\rm LSP}(\omega;\bm{r},\bm{r}')\bigr]
=
-\bm{\nabla}\!\cdot\!\bigl [[\varepsilon (\omega,\bm{r})-1]\bm{\nabla}D_{0}(\omega;\bm{r},\bm{r}')\bigr ].
\end{align}
\end{widetext}
Consider first the lossless case and assume $\varepsilon''=0$ for now. For real dielectric function $\varepsilon (\omega,\bm{r})$, we can expand 
$D_{\rm LSP}$ over the eigenmodes of Eq.~(\ref{app-gauss-law}) as 
%
%
\begin{equation}
\label{app-green-exp}
D_{\rm LSP}(\omega;\bm{r},\bm{r}')=\sum_{n}D_{n}(\omega)\Phi_{n}(\bm{r})\Phi_{n}(\bm{r}'),
\end{equation}
with real coefficients $D_{n}(\omega)$. To solve Eq.~(\ref{app-gauss-green-plas}), let us establish the following useful relations. First, applying the Laplace operator $\Delta'$ (acting on coordinate on $\bm{r}'$) to the LSP Green function expansion (\ref{app-green-exp}) and integrating the result over $V'$ with the factor $\Phi_{n}(\bm{r}')$, we obtain after integrating by parts and using the modes orthogonality:
\begin{equation}
\label{app-lhs}
\int\! dV'\Phi_{n}(\bm{r}')\Delta'D_{\rm LSP}(\omega;\bm{r},\bm{r}')= - D_{n} \Phi_{n}(\bm{r}) \!\int\! dV \bm{E}_{n}^{2}.
\end{equation}
On the other hand, using the same procedure for free-space  Green function $D_{0}(\bm{r}-\bm{r}')$, we obtain
\begin{equation}
\label{app-rhs}
\int\! dV'\Phi_{n}(\bm{r}')\Delta'D_{0}(\bm{r}-\bm{r}')= 4\pi\Phi_{n}(\bm{r}).
\end{equation}
We now use the above relations for finding $D_{n}$. Applying the operator $\Delta'$ to both sides of Eq.~(\ref{app-gauss-green-plas}),  integrating the result over $V'$ with the factor $\Phi_{n}(\bm{r}')$, and using Eqs.~(\ref{app-lhs}) and (\ref{app-rhs}), we obtain
%
%
\begin{equation}
\label{app-gauss-green-plas2}
D_{n} \bm{\nabla}\!\cdot\!\bigl[\varepsilon (\omega,\bm{r})\bm{E}_{n}(\bm{r})\bigr]\int\! dV \!\bm{E}_{n}^{2}=  
4\pi \bm{\nabla}\!\cdot\!\bigl [[\varepsilon (\omega,\bm{r})-1]\bm{E}_{n}(\bm{r})\bigr ].
\end{equation}
Multiplying Eq.~(\ref{app-gauss-green-plas2}) by $\Phi_{n}(\bm{r})$ and integrating  the result over the system volume,  we obtain
\begin{equation}
\label{app-mode-coeff-eq}
\left [D_{n}(\omega)-\dfrac{4\pi}{\int\! dV \bm{E}_{n}^{2}(\bm{r})}\right ]\int\! dV \varepsilon (\omega,\bm{r})\bm{E}_{n}^{2}(\bm{r})=-4\pi.
\end{equation}
The solution of Eq.~(\ref{app-mode-coeff-eq}) for coefficients $D_{n}$ is 
\begin{equation}
\label{app-mode-coeff}
D_{n}(\omega)= 
\dfrac{4\pi}{\int\! dV \bm{E}_{n}^{2}(\bm{r})} 
-\dfrac{4\pi}{\int\! dV \varepsilon (\omega,\bm{r})\bm{E}_{n}^{2}(\bm{r})},
\end{equation}
where the first  term ensures the  condition $D_{n}=0$ at $\varepsilon=1$. Accordingly, the LSP dyadic Green function for the electric fields takes the form 
%
\begin{equation}
\label{app-lsp-green-gen}
\bm{D}_{\rm LSP} (\omega;\bm{r},\bm{r}')=\sum_{n}D_{n}(\omega)\bm{E}_{n}(\bm{r})\bm{E}_{n}(\bm{r}').
\end{equation}

The coefficients (\ref{app-mode-coeff}), which define the Green function  (\ref{app-lsp-green-gen}), are obtained for lossless dielectric function [i.e., with $\varepsilon'' (\omega,\bm{r})=0$] in terms of real eigenmodes $\bm{E}_{n}$ of the Gauss equation Eq.~(\ref{app-gauss-law}). However, as we show below, the expression (\ref{app-mode-coeff}) is valid for the \textit{complex} dielectric function as well. Indeed, let us include $\varepsilon''$  in the derivation of Eq.~(\ref{app-mode-coeff-eq}), which now defines the complex coefficients $\tilde{D}_{n}$ as
\begin{equation}
\label{app-mode-coeff-eq-compl}
\left [\tilde{D}_{n}(\omega)-\dfrac{4\pi}{\int\! dV \tilde{\bm{E}}_{n}^{2}(\bm{r})}\right ]\int\! dV \varepsilon (\omega,\bm{r})\tilde{\bm{E}}_{n}^{2}(\bm{r})=-4\pi,
\end{equation}
where $\varepsilon(\omega,\bm{r})=\varepsilon' (\omega,\bm{r})+i\varepsilon'' (\omega,\bm{r})$ is the complex dielectric function and the complex eigenmodes $\tilde{\bm{E}}_{n}$ constitute the corresponding basis set. Note that for complex $\varepsilon(\omega,\bm{r})$, the complex eigenmodes of Gauss equation (\ref{app-gauss-law}), with $\varepsilon'$ replaced by $\varepsilon$, satisfy the orthogonality relation $\int\! dV_{i} \tilde{\bm{E}}_{n}(\bm{r}) \cdot \tilde{\bm{E}}_{n'}(\bm{r})=\delta_{nn'}\int\! dV_{i} \tilde{\bm{E}}_{n}^{2}(\bm{r})$, implying that the normalization integral in the right-hand side is complex as well. However, within the quasistatic approach, this difficulty can be resolved by treating $\varepsilon'' (\omega,\bm{r})$ as a perturbation. Namely, we adopt the standard perturbation theory  by expanding the new basis set over the unperturbed eigenmodes as $\tilde{\bm{E}}_{n}=\sum_{m}c_{nm}\bm{E}_{m}$ with complex coefficients $c_{nm}$. In the first order, the eigenmodes are unchanged except for the normalization factor, i.e., $\tilde{\bm{E}}_{n}\approx c_{nn}\bm{E}_{n}$, so that $\int\! dV \varepsilon (\omega,\bm{r})\tilde{\bm{E}}_{n}^{2}(\bm{r})=c_{nn}^{2}\int\! dV \varepsilon (\omega,\bm{r})\bm{E}_{n}^{2}(\bm{r})$. We now observe that the higher-order terms of perturbation expansion include non-diagonal terms of the form $\int \! dV_{\rm m}\bm{E}_{n}(\bm{r})\bm{E}_{n'}(\bm{r})$ (for $n\neq n'$), which \textit{vanish} due to the modes' orthogonality property (\ref{app-mode-orth}). Therefore, in \textit{all} orders of the perturbation theory, the complex coefficients $\tilde{D}_{n}$ have the form
\begin{equation}
\label{app-mode-coeff-compl}
\tilde{D}_{n}(\omega)= \frac{1}{c_{nn}^{2}}\left [
\dfrac{4\pi}{\int\! dV \bm{E}_{n}^{2}(\bm{r})} 
-\dfrac{4\pi}{\int\! dV \varepsilon (\omega,\bm{r})\bm{E}_{n}^{2}(\bm{r})}\right ].
\end{equation}
Accordingly, the \textit{exact} quasistatic Green function for \textit{complex} dielectric function $\varepsilon(\omega,\bm{r})=\varepsilon' (\omega,\bm{r})+i\varepsilon'' (\omega,\bm{r})$ takes the form
\begin{widetext}
\begin{align}
\label{app-lsp-green-compl}
\tilde{\bm{D}}_{\rm LSP} 
(\omega;\bm{r},\bm{r}')
=\sum_{n}\tilde{D}_{n}(\omega)\tilde{\bm{E}}_{n}(\bm{r})\tilde{\bm{E}}_{n}(\bm{r}')
= \sum_{n}\left [\frac{4\pi}{\int\! dV \bm{E}_{n}^{2}(\bm{r})} -\sum_{n}\frac{4\pi}{\int\! dV \varepsilon (\omega,\bm{r})\bm{E}_{n}^{2}(\bm{r})}\right ]\bm{E}_{n}(\bm{r})\bm{E}_{n}(\bm{r}'),
\end{align}
\end{widetext}
where the normalization constants $c_{nn}$ cancel out between the numerator and denominator for each mode. Hereafter, we  use the notation $\bm{D}_{\rm LSP} (\omega;\bm{r},\bm{r}')$ for the complex LSP Green function as well.

We now  note that, in the quasistatic regime, the  frequency and coordinate dependencies in the LSP Green's function can be separated out. Using the Gauss equation (\ref{app-gauss-law}) in the integral form $\int\! dV \varepsilon' (\omega_{n},\bm{r})\bm{E}_{n}^{2}(\bm{r})=0$, the volume  integral in Eq.~(\ref{app-mode-coeff})  can be presented as 
\begin{align}
\label{app-int1}
\int\! dV \varepsilon (\omega,\bm{r})\bm{E}_{n}^{2}(\bm{r})
=\left [\varepsilon (\omega)-\varepsilon' (\omega_{n})\right ] \int\! dV_{\rm m}\bm{E}_{n}^{2}(\bm{r}),
\end{align}
where integration in the right-hand side is  carried over the \textit{metal} volume $V_{\rm m}$, while  the dielectric regions' contributions,  characterized by constant permittivities,  cancel each other out. In a similar way, we have 
%
\begin{equation}
\label{app-int2}
\int\! dV \bm{E}_{n}^{2}(\bm{r})=\left [1-\varepsilon' (\omega_{n})\right ] \int\! dV_{\rm m}\bm{E}_{n}^{2}(\bm{r}),
\end{equation}
and the LSP Green's function takes the form
\begin{widetext}
\begin{align}
\label{app-lsp-green}
\bm{D}_{\rm LSP} (\omega;\bm{r},\bm{r}')
=\sum_{n}\dfrac{\bm{E}_{n}(\bm{r})\bm{E}_{n}(\bm{r}')}{\int\! dV_{\rm m} \bm{E}_{n}^{2}}
\left [\frac{4\pi}{1-\varepsilon' (\omega_{n})}-\frac{4\pi}{\varepsilon (\omega)-\varepsilon' (\omega_{n})}\right ].
\end{align}
Using relation (\ref{app-int2}), the Green function can be recast as
\begin{align}
\label{app-lsp-green2}
\bm{D}_{\rm LSP} (\omega;\bm{r},\bm{r}') 
=
\sum_{n}\dfrac{\bm{E}_{n}(\bm{r})\bm{E}_{n}(\bm{r}')}{\int\! dV_{\rm m} \bm{E}_{n}^{2}}
\left [\frac{4\pi}{1-\varepsilon' (\omega_{n})}\frac{\varepsilon (\omega)-1}{\varepsilon (\omega)-\varepsilon' (\omega_{n})}\right ]
=4\pi\sum_{n}
\dfrac{\bm{E}_{n}(\bm{r})\bm{E}_{n}(\bm{r}')}{\int\! dV\bm{E}_{n}^{2}}\frac{\varepsilon (\omega)-1}{\varepsilon (\omega)-\varepsilon' (\omega_{n})},
\end{align}
\end{widetext}
where the volume integral in the second equation extends over entire space.

The full near field Green dyadic  $\bm{D}(\omega;\bm{r},\bm{r}') =\bm{D}_{0} (\bm{r}-\bm{r}') +\bm{D}_{\rm LSP} (\omega;\bm{r},\bm{r}')$ includes direct dipole term and LSP contribution. In the case when $\bm{r}$ is outside the metal while $\bm{r}'$ is inside of it, the direct dipole term, given by Eq.~(\ref{app-green-free-expansion}) and first term in Eq.~(\ref{app-lsp-green}) \textit{cancel} each other out and we obtain
\begin{widetext}
\begin{align}
\label{app-green-full-in-out}
\bm{D}(\omega;\bm{r},\bm{r}')
=\bm{D}_{0} (\bm{r}-\bm{r}') +\bm{D}_{\rm LSP} (\omega;\bm{r},\bm{r}')
=\sum_{n}\dfrac{\bm{E}_{n}(\bm{r})\bm{E}_{n}(\bm{r}')}{\int\! dV_{\rm m} \bm{E}_{n}^{2}} \frac{4\pi}{\varepsilon' (\omega_{n})-\varepsilon (\omega)},
\end{align}
\end{widetext}
indicating that, in this case, only the LSP resonances contribute.

In the case when both $\bm{r}$ and $\bm{r}'$ are in the same region of a metal-dielectric structure, e.g., inside the metal, no such cancellation takes place for the dyadic Green function $\bm{D}(\omega;\bm{r},\bm{r}')$. However,   using Eqs.~(\ref{app-lsp-green}) and (\ref{app-relation-in-out}), the following integral relation holds for\textit{any} $\bm{r}$,
\begin{equation}
\label{app-green-integral}
\int\! dV_{\rm m}\bm{D}(\omega;\bm{r},\bm{r}_{\rm m})\bm{E}_{n}(\bm{r}_{\rm m})
=\frac{4\pi\bm{E}_{n}(\bm{r})}{\varepsilon' (\omega_{n})-\varepsilon (\omega)},
\end{equation}
where the contributions from $\bm{D}_{0} $ and first term in Eq.~(\ref{app-lsp-green}) cancel each other out.

\subsection{The Lorentzian limit}
In the \textit{Lorentzian} approximation, the dielectric function $\varepsilon(\omega)$ in Eq.~(\ref{app-lsp-green}) is expanded near the LSP frequencies $\omega_{n}$ as 
\begin{equation}
\label{app-expand}
\varepsilon(\omega)-\varepsilon' (\omega_{n})=(\omega-\omega_{n})\varepsilon'_{n}+ i\varepsilon''(\omega_{n}),
\end{equation}
where we denoted $\varepsilon'_{n}\equiv \partial\varepsilon' (\omega_{n})/\partial \omega_{n}$. The Lorentzian LSP Green function has the form \cite{shahbazyan-prl16,shahbazyan-prb18}
\begin{equation}
\label{app-green-lorentzian}
\bm{D}_{\rm LSP}^{L} (\omega;\bm{r},\bm{r}')=\frac{1}{\hbar}\sum_{n}\frac{\tilde{\bm{E}}_{n}(\bm{r})\tilde{\bm{E}}_{n}(\bm{r}')}{\omega_{n}-\omega-i\gamma_{n}/2},
\end{equation}
where 
\begin{equation}
\label{app-lsp-field-norm}
\tilde{\bm{E}}_{n}(\bm{r})
=\sqrt{\frac{4\pi\hbar}{\varepsilon'_{n}}}
\dfrac{\bm{E}_{n}(\bm{r})}{\left (\int\! dV_{\rm m} \bm{E}_{n}^{2}\right )^{1/2}},
\end{equation}
are normalized LSP mode fields introduced to match the standard Lorentzian expression for the Green function and $\gamma_{n}=2\varepsilon''(\omega_{n})/\varepsilon'_{n}$ is the LSP decay rate. In terms of normalized fields, the LSP optical dipole moment is defined as $\bm{\mu}_{n}=\!\int\! dV\chi' (\omega_{n},\bm{r})\tilde{\bm{E}}_{n}(\bm{r})$, where  $\chi =(\varepsilon-1)/4\pi$ is susceptibility, and the LSP radiative decay rate has the standard form $\gamma_{n}^{\rm rad}=4\mu_{n}^{2}\omega_{n}^{3}/3\hbar c^{3}$.


\section{Derivation of optical polarizability for metal nanoparticles of arbitrary shape}

In the following, we consider binary systems, i.e., metal NPs in a dielectric medium with permittivity $\varepsilon_{d}$, which we set $\varepsilon_{d}=1$ for now.  In the presence of incident field $\bm{E}_{0}e^{-i\omega t}$ that is uniform at the system scale, the electric field $\bm{E}(\bm{r})$ at any point $\bm{r}$ inside or outside the metal is   
%
\begin{align}
\label{app-field-at-r}
\bm{E}(\bm{r})=\bm{E}_{0}+\chi(\omega)\int dV_{\rm m}\bm{D}(\omega;\bm{r},\bm{r}_{\rm m})\bm{E}_{0},
\end{align}
where $\bm{D}(\omega;\bm{r},\bm{r}')=\bm{D}_{0}(\bm{r},\bm{r}')+\bm{D}_{\rm LSP}(\omega;\bm{r},\bm{r}')$ is near-field dyadic Green function with LSP contribution  $\bm{D}_{\rm LSP}(\omega;\bm{r},\bm{r}')$ given by Eq.~(\ref{app-lsp-green}). Since, inside the metal, the LSP modes $\bm{E}_{n}(\bm{r})$ are regular functions of $\bm{r}$, the external field in the integrand of Eq.~(\ref{app-field-at-r}) can be expanded as 
\begin{equation}
\label{app-e0-expand}
\bm{E}_{0}=\sum_{n}c_{n0}\bm{E}_{n}(\bm{r}_{\rm m}),
~~~
c_{n0}=\frac{\int dV_{\rm m}\bm{E}_{n}(\bm{r}_{\rm m})\!\cdot\! \bm{E}_{0}}{\int dV_{\rm m}\bm{E}_{n}^{2}(\bm{r}_{\rm m})},
\end{equation}
and, upon using Eq.~(\ref{app-green-integral}), we obtain
\begin{align}
\label{app-field-at-r3}
\bm{E}(\bm{r})=\bm{E}_{0}-\sum_{n}c_{n0}
\bm{E}_{n}(\bm{r})\tilde{\alpha}_{n}(\omega)
\end{align}
where
\begin{equation}
\tilde{\alpha}_{n}(\omega)=\frac{\varepsilon(\omega)-1}{\varepsilon(\omega)-\varepsilon'(\omega_{n})}
\end{equation}
is \textit{dimensionless} $n$-polarizability of NP. Note that if the external field excites  dipolar modes, e.g., longitudinal or transverse modes in a nanorod, only those modes contribute to the sum in Eq.~(\ref{app-field-at-r3}) as for higher-order modes the coefficients $c_{n0}$ vanish. For $\bm{r}$ inside the metal, it can be further simplified by using expansion (\ref{app-e0-expand}) for the first term, and we obtain
\begin{align}
\label{app-field-in}
\bm{E}(\bm{r}_{\rm m})=\sum_{n}c_{n0}
\bm{E}_{n}(\bm{r}_{\rm m})
\frac{1-\varepsilon'(\omega_{n})}{\varepsilon(\omega)-\varepsilon'(\omega_{n})}.
\end{align}
We now define the induced LSP dipole moment as  $\bm{p}=\chi(\omega) \int dV_{\rm m}\bm{E}(\omega,\bm{r})$,  and using Eq.~(\ref{app-field-in}), obtain
 $\bm{p}(\omega)=\sum_{n}\bm{p}_{n}(\omega)$, where  $\bm{p}_{n}(\omega)=\bm{\alpha}_{n}(\omega)\bm{E}_{0}$ is induced dipole moment of the $n$th LSP mode and $\bm{\alpha}_{n}(\omega)$ is its polarizability tensor. The latter is obtained as 
\begin{equation}
\bm{\alpha}_{n}(\omega)=\alpha_{n}(\omega)\bm{e}_{n}\bm{e}_{n}=V_{n}\tilde{\alpha}_{n}(\omega)\bm{e}_{n}\bm{e}_{n},
\end{equation}
where $\alpha_{n}(\omega)=V_{n}\tilde{\alpha}_{n}(\omega)$ is the NP $n$-polarizability, $\bm{e}_{n}=\int\! dV_{\rm m}\bm{E}_{n}/|\int\! dV_{\rm m}\bm{E}_{n}|$ is unit vector for  LSP mode polarization and $V_{n}$ is \textit{effective} system volume  defined as
\begin{equation}
\label{app-Vn}
V_{n}=s_{n}V_{\rm m}\frac{1-\varepsilon'(\omega_{n})}{4\pi},
~~~
s_{n}=\dfrac{\left (\int\! dV_{\rm m}\bm{E}_{n}\right )^{2}}{V_{\rm m}\int\! dV_{\rm m} \bm{E}_{n}^{2}}.
\end{equation}
Here,  $s_{n}\leq 1$ is parameter depending on system geometry which is $s_{n}=1$ for spherical and spheroidal NPs (see below). Note that for a spherical NP of radius $a$, we have $\varepsilon'(\omega_{n})=-2$ and hence $V_{n}=a^{3}$, recovering the standard expression for its polarizability.
%
For larger NPs beyond the quasistatic limit, the LSP radiation damping is included in the standard way via replacement $ \alpha_{n} \rightarrow \alpha_{n}\left [ 1- (2i/3)k^{3}\alpha_{n}\right]^{-1}$. Restoring the medium dielectric constant $\varepsilon_{d}$, we finally obtain
 \begin{equation}
 \label{app-pol-small2}
\alpha_{n}(\omega)=\dfrac{V_{n}[\varepsilon(\omega)-\varepsilon_{d}]}{\varepsilon(\omega)\!-\!\varepsilon'(\omega_{n})\!-\!\frac{2i}{3}k^{3}V_{n}[\varepsilon(\omega)\!-\!\varepsilon_{d}]},
 \end{equation}
where $V_{n}=V_{\rm m}|\varepsilon'(\omega_{n})/\varepsilon_{d}-1|s_{n}/4\pi$. In the Lorentzian approximation, expanding $\varepsilon(\omega)$ near $\omega_{n}$ according Eq.~(\ref{app-expand}) and using again Eq.~(\ref{app-lsp-field-norm}), we recover the standard expression for polarizability tensor of LSP treated as a localized dipole
\begin{equation}
 \label{app-pol-L-small2}
\bm{\alpha}_{n}^{L}(\omega)=\frac{1}{\hbar}\frac{\bm{\mu}_{n}\bm{\mu}_{n}}{\omega_{n}-\omega-i\gamma_{n}/2},
\end{equation}
where the LSP decay rate now includes both non-radiative and radiative processes: $\gamma_{n}=2\varepsilon'' (\omega_{n})/\varepsilon'_{n}+4\mu_{n}^{2}\omega_{n}^{3}/3\hbar c^{3}$. 

\section{Polarizability of spheroidal particles}

Here we demonstrate that our approach reproduces accurately the results for polarizability of nanostructures which are known analytically. Specifically, we consider  the longitudinal mode of a small prolate spheroidal particle with semi-major axis $a$ and semi-minor axis $b$. The particle polarizability, which includes the radiation damping, is
\begin{equation}
\label{pol-rad-spheroid}
\alpha=\frac{V_{m}}{4\pi}\dfrac{\varepsilon (\omega)-1}{1+[\varepsilon (\omega)-1]\left (L-i\dfrac{4\pi^{2}}{3\lambda^{3}}V_{m}\right )},
\end{equation}
where $V_{\rm m}=4\pi ab^{2}/3$ is the particle volume, $\lambda=2\pi c/\omega$ is the wavelength, and $L$ is the depolarization factor given by
\begin{equation}
L=\frac{1-e^{2}}{e^{2}}\left [\frac{1}{2e}\ln\frac{1+e}{1-e}-1\right ],
\end{equation}
with $e=\sqrt{1-b^{2}/a^{2}}$. Note that $L=1/3$ for a spherical particle (i.e., for $a=b$). Let us show that the same result follows from our general expression for polarizability, which, for a single mode, has the form

 \begin{equation}
 \label{pol-rad-gen}
\alpha_{n}(\omega)
 =V_{n}\, \dfrac{\varepsilon(\omega)-1}{\varepsilon(\omega)-\varepsilon'(\omega_{n})-\dfrac{2i\omega^{3}}{3c^{3}}V_{n}[\varepsilon(\omega)-1]},
 \end{equation}
where $\omega_{n}$ is the longitudinal mode frequency and  $V_{n}=V_{\rm m}|\varepsilon'(\omega_{n})-1|s_{n}/4\pi$ is the corresponding effective volume. Note that Eq.~(\ref{pol-rad-gen}) can be recast in the form [compare to Eq.~(\ref{pol-rad-spheroid})]
\begin{equation}
\alpha_{n}=\frac{s_{n}V_{m}}{4\pi}\dfrac{\varepsilon (\omega)-1}{1+[\varepsilon (\omega)-1]\left (L_{n}-i\dfrac{4\pi^{2}}{3\lambda^{3}}s_{n}V_{m}\right )},
\end{equation}
where $L_{n}=|\varepsilon'(\omega_{n})-1|^{-1}$. Below we show  that $L_{n}=L$ and $s_{n}=1$, which insures that $\alpha_{n}=\alpha$.

\subsection{Eigenmodes for a spheroidal nanoparticle}

Consider a prolate spheroid with semiaxis $a$ along the symmetry axis and semiaxis  $b$ in the symmetry plane ($a>b$). We use standard notations for spheroidal coordinates ($\xi,\eta,\phi$) where $\xi$ is the "radial" coordinate while $\eta=\cos \theta$ and $\phi$ parametrize the surface.  The scaling factors are given by
\begin{align}
h_{\xi}=f\sqrt{\frac{\xi^{2}-\eta^{2}}{\xi^{2}-1}},
~~
h_{\eta}=f\sqrt{\frac{\xi^{2}-\eta^{2}}{1-\eta^{2}}},
\nonumber\\
~~
h_{\phi}=f\sqrt{(\xi^{2}-1)(1-\eta^{2})},
\end{align}
where $f =\sqrt{a^{2}-b ^{2}}$ is half distance between the foci, and spheroid surface corresponds to $\xi_{0}=a/f$. The spheroidal and Cartesian coordinates are related as
\begin{align}
x=h_{\phi}\cos\phi,~~
y=h_{\phi}\sin\phi,~~
z=f\eta\xi.
\end{align}
The gradient operator has the form 
\begin{equation}
\bm{\nabla}=\hat{\bm{\xi} }h_{\xi}^{-1}\partial/\partial\xi+\hat{\bm{\eta} }h_{\eta}^{-1}\partial/\partial\eta+\hat{\bm{\phi} }h_{\phi}^{-1}\partial/\partial\phi,
\end{equation}
%
where $\hat{\bm{\xi} }$, $\hat{\bm{\eta}}$ and $\hat{\bm{\phi} }$ are spheroidal unit vectors,
\begin{align}
\label{unit-spheroid}
&\hat{\bm{\xi} }=\frac{1}{h_{\xi}}\frac{\partial \bm{r}}{\partial \xi}=
\frac{f\xi}{h_{\eta}}
\left (\cos\phi \hat{\bm{x}} + \sin\phi \hat{\bm{y}}\right )
+ \frac{f\eta}{h_{\xi}}\hat{\bm{z}},
\nonumber\\
&\hat{\bm{\eta} }=\frac{1}{h_{\eta}}\frac{\partial \bm{r}}{\partial \eta}=
-\frac{f\eta}{h_{\xi}}
\left (\cos\phi \hat{\bm{x}} + \sin\phi \hat{\bm{y}}\right )
+ \frac{f\xi}{h_{\eta}}\hat{\bm{z}},
\nonumber\\
&\hat{\bm{\phi} }=\frac{1}{h_{\phi}}\frac{\partial \bm{r}}{\partial \phi}=
-\sin\phi \hat{\bm{x}} + \cos\phi \hat{\bm{y}}.
\end{align}
The volume and surface elements  are, respectively, $dV=h_{\xi}h_{\eta}h_{\phi} d\xi d\eta d\phi$ and $dS=h_{\eta}h_{\phi} d\eta d\phi$, while the full surface area  and volume are
\begin{equation}
S=2\pi b^{2}\left (1+\frac{2\alpha}{\sin 2\alpha}\right ),
~~
V= \frac{4\pi}{3}\, a b^{2},
\end{equation}
where $\alpha=\arccos (b/a)$.  For $b/a\ll 1$, we have $S=\pi^{2} ab$.


Let us turn to the eigenmodes. The potential $\Phi_{n}$ for a longitudinal dipole mode is 
%
\begin{equation}
\label{potentials}
\Phi_{n}=fR_{n}(\xi)P_{1}(\eta),
\end{equation}
with radial field components given by
\begin{align}
\label{R-long}
&R_{n}(\xi)=P_{1}(\xi),~~~\text{for $\xi<\xi_{0}$},
\nonumber\\
&R_{n}(\xi)=Q_{1}(\xi)P_{1}(\xi_{0})/Q_{1}(\xi_{0}),~~~\text{for $\xi>\xi_{0}$}.
\end{align}
where $P_{l}(x)$ and $Q_{l}(x)$ are the Legendre functions. The electric field has the form
\begin{equation}
\label{field-spheroid}
\bm{E}_{n}=-\bm{\nabla}\Phi_{n}=-\frac{f\eta}{h_{\xi}} R'_{n}(\xi)\hat{\bm{\xi}}-\frac{f}{h_{\eta}} R_{n}(\xi)\hat{\bm{\eta}},
\end{equation}
where $\hat{\bm{\xi}}$ and $\hat{\bm{\eta}}$ are given by Eq.~(\ref{unit-spheroid}), and prime stands for derivative. 
The  relevant Legendre functions and their derivatives have the form
\begin{align}
\label{legendre}
P_{1}(\xi)=\xi, ~~~ Q_{1}(\xi)=\frac{\xi}{2}\ln\left [\frac{\xi + 1}{\xi - 1}\right ] - 1,
~~~
\nonumber\\
Q'_{1}(\xi)=\frac{1}{2}\ln\left [\frac{\xi + 1}{\xi - 1}\right ] - \frac{\xi}{\xi^2 - 1}.
\end{align}

\subsection{Evaluation of $s_{n}$ and $L_{n}$}

The parameter $s_{n}$ is defined as
\begin{align}
\label{sn}
s_{n}=\dfrac{\left (\int\! dV_{\rm m}\bm{E}_{n}\right )^{2}}{V_{\rm m}\int\! dV_{\rm m} \bm{E}_{n}^{2}},
\end{align}
where the integrals are taken over the  spheroid volume. Note that $s_{n}$ is independent of the overall field normalization. Starting with the denominator, using Eq.~(\ref{field-spheroid}) with $R_{n}(\xi)=\xi$, we have $\bm{E}_{n}^{2}=1$, yielding $\int\! dV_{\rm m} \bm{E}_{n}^{2}=V_{\rm m}$. Turning to the numerator of Eq.~(\ref{sn}), we note that $\int\! dV_{\rm m}\bm{E}_{n}
=- \int\! dS \Phi_{n}\hat{\bm{\xi}}$ and,  using Eqs.~(\ref{unit-spheroid}) and (\ref{potentials}), we obtain
\begin{align}
\int\! dS \Phi_{n}\hat{\bm{\xi}}
=2\pi  f^{2} \xi_{0}\int_{-1}^{1} d\eta \eta^{2}\,\frac{h_{\phi}h_{\eta}}{h_{\xi}} \hat{\bm{z}}
~~~~~~~~~
\nonumber\\
=\frac{4\pi}{3}f^{3}\xi_{0}(\xi_{0}^{2}-1)\hat{\bm{z}}
=\frac{4\pi}{3}ab^{2}\hat{\bm{z}}
=V_{\rm m}\hat{\bm{z}},
\end{align}
leading to  $s_{n}=1$.

Turning to $L_{n}$, the mode frequency $\omega_{n}$ follows from the Gauss equation (\ref{gauss-law}) by matching the normal components of the electric field (\ref{field-spheroid}) across the interface:
\begin{equation}
\varepsilon' (\omega_{n})=\xi_{0}Q'_{1}(\xi_{0}) /Q_{1}(\xi_{0}).
\end{equation}
Using the explicit expressions (\ref{legendre}), we obtain
\begin{widetext}
\begin{align}
L_{n}=\frac{1}{|\varepsilon' (\omega_{n})-1|}= \frac{Q_{1}(\xi_{0})}{Q_{1}(\xi_{0})-\xi_{0}Q'_{1}(\xi_{0})} 
= (\xi_{0}^{2}-1)\left [   \frac{\xi_{0}}{2}\ln\left (\frac{\xi_{0}+ 1}{\xi_{0} - 1}\right ) - 1    \right ].
\end{align}
\end{widetext}
Finally, noting that $\xi_{0}=a/\sqrt{a^{2}-b^{2}}=1/e$, we obtain $L_{n}=L$, which proves the equivalence of polarizabilities given by Eqs. (\ref{pol-rad-spheroid}) and (\ref{pol-rad-gen}).


\end{document}